# Imbalanced Multi-label Classification for Business-related Text with Moderately Large Label Spaces

Muhammad Arslan[1][0000-0003-3682-7002] and Christophe Cruz[1][0000-0002-5611-9479]

[1,2] Laboratoire Interdisciplinaire Carnot de Bourgogne (ICB), Université de Bourgogne,
9 Av. Alain Savary, 21000 Dijon, France
[1]`muhammad.arslan@u-bourgogne.fr`

**Abstract.** In this study, we compared the performance of four different methods for multi-label text classification using a specific imbalanced business dataset. The four methods we evaluated were fine-tuned BERT, Binary Relevance, Classifier Chains, and Label Powerset. The results show that fine-tuned BERT outperforms the other three methods by a significant margin, achieving high values of accuracy, F1-Score, Precision, and Recall. Binary Relevance also performs well on this dataset, while Classifier Chains and Label Powerset demonstrate relatively poor performance. These findings highlight the effectiveness of fine-tuned BERT for multi-label text classification tasks, and suggest that it may be a useful tool for businesses seeking to analyze complex and multi-faceted texts.

**Keywords:** Business domain, Multi-lingual, Performance comparison, News documents, Fine-tuned BERT

## 1 Introduction

In today's digital age, businesses are generating vast amounts of data through various channels, such as online news articles, press releases, and company websites [1]. This data contains valuable information regarding the launch of new products, services, and business initiatives, among other things. However, analyzing this data manually can be a tedious and error-prone process, especially when it comes to identifying key insights and trends, especially in cases of moderately large label spaces.

Multi-label text classification is the task of assigning one or more labels to a text document [2]. Multilabel text classification using BERT [3] and Problem Transformation approaches [4] can greatly enhance the efficiency and accuracy of analyzing textual business data, including moderately large label spaces. These machine learning models are designed to process large amounts of text data and can automatically identify key topics and themes present in the text. With the ability to classify text into multiple categories, these models can handle more complex datasets and provide a more nuanced understanding of the information contained within.

For instance, a business seeking to understand customer feedback on a new product can use multilabel text classification to analyze reviews and identify which aspects of the product are most highly praised or criticized. Similarly, a company seeking to expand its operations to new geographic regions can analyze news articles and identify the sentiment towards the company in different regions.

In this study, we have compared the performance of four different methods for multi-label text classification using a specific imbalanced business-related dataset. Imbalanced dataset refers to a dataset in which the distribution of text examples among the different categories are unequal. The first method, Fine-tuned BERT, is a popular pre-trained language model that has been fine-tuned for a specific task. The other three methods evaluated in the study were Binary Relevance, Classifier Chains, and Label Powerset, which are all traditional multi-label classification methods.

The structure of this paper is as follows. In Section 2, we provide a review of the background of the multi-label text classification approaches used in this article. Section 3 introduces the proposed work. In Section 4, we discuss the results and implications of our approach. Finally, we conclude the paper in Section 5.

## 2      Background

Problem Transformation approaches are a family of techniques used in multi-label classification tasks to transform the original multi-label problem into one or more simpler, more manageable classification tasks [5]. These approaches are often used when there are a large number of possible labels, which can make the multi-label classification problem computationally expensive and challenging. The most common Problem Transformation approaches include Binary Relevance, Classifier Chains, and Label Powerset [5].

In Binary Relevance method [6], a separate binary classifier is trained for each label, and each classifier predicts whether or not the input belongs to that particular label. The main advantage of the Binary Relevance method is its simplicity and flexibility. It can work with any binary classifier, and the classifiers can be trained independently, making it easy to add or remove labels without affecting the performance of other classifiers. However, the method does not consider any correlations between the labels, which may affect the overall accuracy of the multilabel classification task.

The Classifier Chain method [7] use a chain of binary classifiers to predict the labels. In this method, the labels are treated as a sequence, and the classifiers are trained in the order of the label sequence. The main advantage of the Classifier Chain method is its ability to model the correlations between labels, which can lead to improved accuracy in the multilabel classification task. However, the method can be computationally expensive, especially if there are many labels in the dataset.

The Label Powerset method [8] involves transforming the multilabel problem into a multiclass problem. In this method, each unique combination of labels is treated as a separate class, and a multiclass classifier is trained to predict the class for each input. The main advantage of the Label Powerset method is its ability to handle any number of labels, and it can capture complex dependencies between labels. However, the method suffers from the curse of dimensionality, as the number of classes grows exponentially with the number of labels in the dataset.

Apart from Problem Transformation approaches, fine-tuning an existing pre-trained BERT (Bidirectional Encoder Representations from Transformers) model for multi-label text classification is a popular and effective approach in the existing literature [9] for imbalanced datasets having large label spaces (i.e. high number of categories i.e. labels). Fine-tuning a pre-trained BERT model for multi-label text classification involves training the model on a specific dataset, with the labels and associated text inputs. During training, the weights of the BERT model are updated to optimize the performance of the model on the specific multi-label text classification task.

There are numerous studies exist covering advanced methods for multi-label classification [10 - 12]. However, their applicability to business-related imbalanced dataset with moderately large label spaces has not been extensively explored in the literature. In order to bridge this gap, this article presents a comparative analysis of four different techniques using a business-related dataset. This study aims to provide valuable insights into the effectiveness of these methods for multi-label classification tasks in a business context, and to inform future research in this area.

## 3  Analyzing models for business-related text

In this section, we will provide an overview of the dataset used for our multi-label classification tasks. We will then use this dataset to train and evaluate several multi-label classification models.

### 3.1  Dataset

In preparation for the analysis of various multi-label classification models, a business-related dataset was obtained from the French company FirstECO. This dataset comprises 28,941 business-related texts in French, each of which can be classified into one or more of 80 possible categories. These categories are basically grouped into seven parent domains, which include Intangible development, activity, products, Material investment, Increased standby, Financial development, Company life, Geographical development, and Public finances. For data confidentiality reasons, only the 1st level of categories (labels) are reported. Each category in the dataset contains at least 50 to 100 text examples. However, the dataset remains imbalanced because the distribution of text examples among the different categories are unequal (see Fig. 1). Prior to analysis, the dataset was subjected to pre-processing procedures.

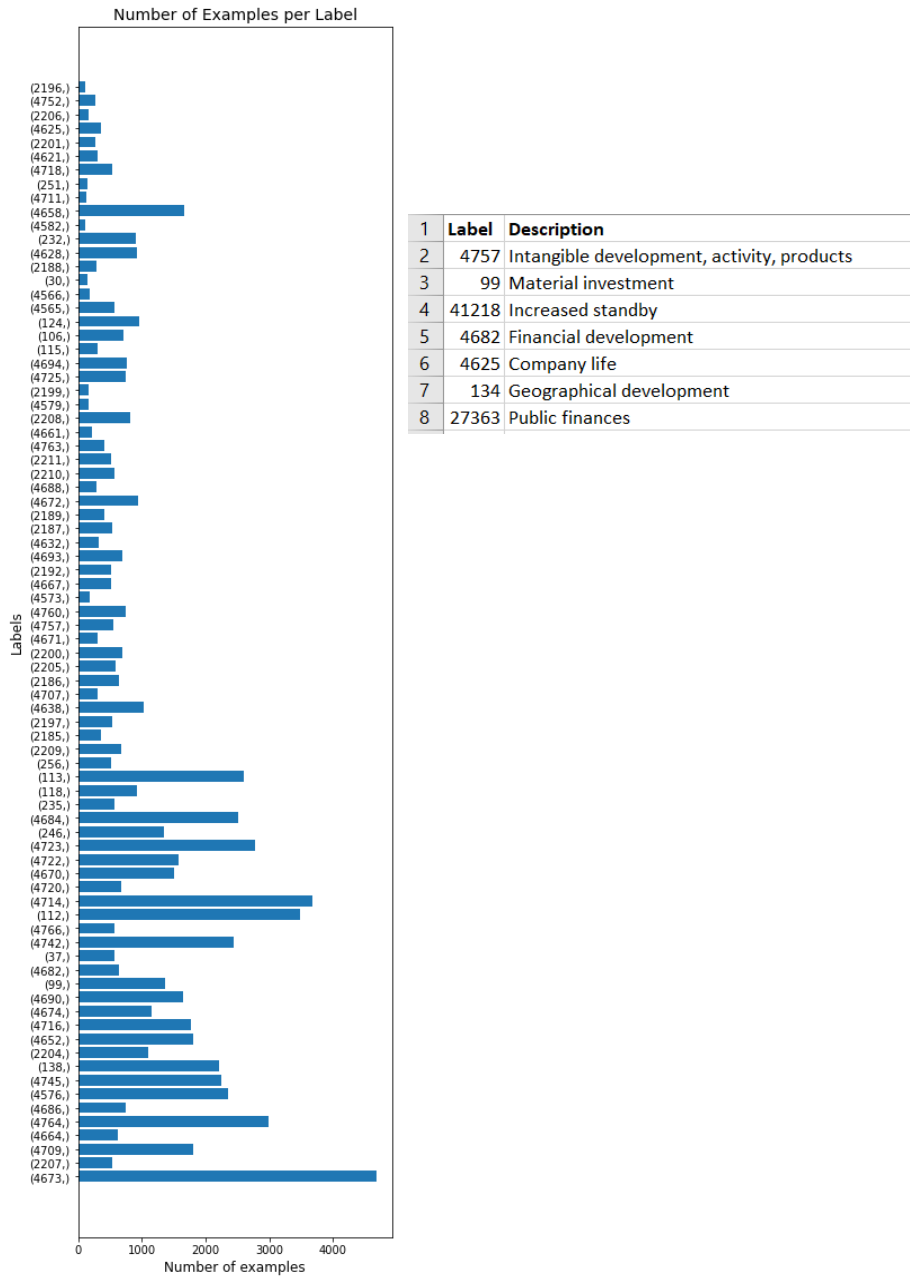

**Fig. 1.** a) Number of examples per label (left), b) Description of each label (right)

The text preprocessing is used as to clean and prepare text data before it is used for text classification. It takes the text data as input and applies several preprocessing steps to each text row. First, it removes punctuation and digits from the text using regular expressions. Then, it converts the text to lowercase and tokenizes it into words using the tokenizer. Next, it removes stop words from the text using a predefined set of stop words. Finally, it joins the preprocessed words back into a string and appends it to a new list. The preprocessing function is a crucial step in preparing text data for machine learning tasks, as it can help to reduce noise and improve the accuracy of the resulting models.

### 3.2  Implementation

The implementation involves the execution of Problem Transformation approaches and fine-tuning BERT model for imbalanced multi-label classification for business-related text with moderately large label spaces.

**a) Problem Transformation approaches:** The process starts by importing necessary modules for data preparation, model training, and evaluation. The imported modules include GaussianNB and MultinomialNB for Naive Bayes classification, and Accuracy_score for evaluation metrics, train_test_split for splitting the data into training and testing sets, and TfidfVectorizer for transforming the text data into feature vectors. The scikit-multilearn library is also imported to support multi-label classification problems. The process then creates an instance of TfidfVectorizer to convert the text data into feature vectors. The TfidfVectorizer is set to use inverse document frequency and normalization.

Next, the process creates an instance of MultiLabelBinarizer and applies it to the labels. The MultiLabelBinarizer transforms the list of labels into a binary matrix where each row corresponds to an instance and each column corresponds to a unique label. Then, the data is split into training and testing sets using train_test_split, with a test size of 20%. Finally, the process returns X_train, X_test, Y_train, and Y_test, which are the feature matrices and label matrices for the training and testing sets, respectively. These matrices are used to train and evaluate multi-label classification models based on Problem Transformation approaches, which are; Binary Relevance, Classifier Chain, and Label Powerset in our case. Finally, the performance of these approaches is evaluated on the testing dataset using various metrics such as accuracy, precision, recall, and F1 score (see Table 1). Accuracy is the percentage of correctly classified samples out of the total number of samples. F1-Score is a weighted average of Precision and Recall, where the F1-Score gives equal importance to both Precision and Recall. Precision measures how precise the model's positive predictions are. Lastly, Recall measures how well the model can identify all positive samples.

**b) Fine-tuning BERT:** To fine-tune existing BERT-based model for text classification, the model "bert-base-multilingual-cased" [13] is chosen as it supports the French language. The process of fine-tuning starts with importing the necessary libraries such as NumPy, Pandas, Scikit-learn, PyTorch, and Transformers. Then, a number of hyperparameters are set, including Max_Len, which is set to 80 and represents the max-imum length of input sequences. Train_Batch_Size is set to 16, and Valid_Batch_Size is set to 8. The process also specifies the number of Epochs to train the model, which is set to 5, and sets the learning rate to 1e-05. Additionally, a pre-trained BERT tokenizer using the BertTokenizer class is used from the transformer's library.

Furthermore, the data is split into training and testing datasets using a Train_Size of 0.8. The training dataset is created by randomly sampling 80% of the data from the original dataset, while the testing dataset is created by dropping the samples in the training dataset from the original dataset. The final training dataset has 23,153 samples, while the testing dataset has 5,788 samples. The BERT model is trained on the training dataset by feeding batches of input sequences to the model, computing the loss, and optimizing the weights using backpropagation. Finally, the model's performance is evaluated on the testing dataset using accuracy, precision, recall, and F1 score (see Table 1).

**Table 1.** Comparative analysis of different multi-label classification approaches

| Parameter | Binary Relevance | Classifier Chains | Label Powerset | Fine-tuned BERT |
|---|---|---|---|---|
| **Accuracy** | 0.730 | 0.103 | 0.143 | 0.895 |
| **F1-Score** | 0.936 | 0.539 | 0.278 | 0.978 |
| **Precision** | 0.952 | 0.590 | 0.350 | 0.948 |
| **Recall** | 0.922 | 0.495 | 0.230 | 0.988 |

Accuracy, F1-score, precision, and recall are commonly used performance metrics that can be used to evaluate the effectiveness of a classifier. Accuracy measures the fraction of instances that are correctly classified by the classifier. Precision measures the fraction of correctly identified positive instances among all instances predicted as positive, while recall measures the fraction of correctly identified positive instances among all positive instances in the data. Using these definitions, we can compute accuracy, precision and recall as follows:

$$Accuracy = (TP + TN) / (TP + TN + FP + FN)$$

$$Precision = TP / (TP + FP)$$

$$Recall = TP / (TP + FN)$$

Where, True positives (TP) are instances that are positive and are correctly classified as positive by the classifier. False positives (FP) are instances that are negative but are incorrectly classified as positive by the classifier. True negatives (TN) are instances that are negative and are correctly classified as negative by the classifier and False negatives (FN) are instances that are positive but are incorrectly classified as negative by the classifier. However, accuracy may not be a suitable metric to use when the classes are imbalanced. This is because a classifier that simply predicts the majority class for all instances would achieve high accuracy even if it performs poorly on the minority class. To address this problem, we can use the F1-Score, which is a harmonic mean of precision and recall. It combines both precision and recall into a single metric that balances the trade-off between them. The F1-Score is defined as:

$$F1\text{-Score} = 2 * (Precision * Recall) / (Precision + Recall)$$

The F1-Score ranges between 0 and 1, with a value of 1 indicating perfect precision and recall. Note that precision measures the accuracy of the positive predictions made by the classifier, while recall measures the completeness of the positive predictions made by the classifier.

## 4    Results

The Table 1 displays the performance of four different methods for multilabel text classification on a dataset with 80 possible labels. The first method, Binary Relevance, achieves an accuracy of 0.730, F1-Score of 0.936, Precision of 0.952, and Recall of 0.922. This method creates a separate binary classifier for each label and assigns a label to each text independently. The second method, Classifier Chains, achieves an accuracy of 0.103, F1-Score of 0.539, Precision of 0.590, and Recall of 0.495. This method builds a chain of classifiers where each classifier considers the predictions of the previous classifiers in the chain. The third method, Label Powerset, achieves an accuracy of 0.143, F1-Score of 0.278, Precision of 0.350, and Recall of 0.230. This method transforms the multilabel classification problem into a multi-class classification problem by assigning each unique combination of labels to a single class. The fourth method, fine-tuned BERT, achieves the highest accuracy of 0.895, F1-Score of 0.978, Precision of 0.948, and Recall of 0.988.

One important factor that has impacted the performance of a multilabel classification model is the imbalanced distribution of labels in the dataset. In other words, some labels may have significantly more instances than others, making it challenging for the model to learn to classify the minority labels correctly. In this context, the provided table suggests that the dataset used to train and evaluate the model is imbalanced, with some labels having significantly fewer instances than others. To address the problem of imbalanced labels, various techniques can be used, such as oversampling or undersampling, class weighting, or ensemble methods. These techniques aim to balance the distribution of labels in the training set, which can improve the model's performance on the minority labels. However, choosing the appropriate technique depends on the specific characteristics of the dataset and the algorithm used. Nonetheless, the discussed model's (i.e. fine-tuned BERT in our case) performance is still competitive, and further investigation may be necessary to identify the factors that affect its performance on different labels.

## 5      Discussion and Conclusion

The study focused on the problem of imbalanced multi-label classification for business-related text with moderately large label spaces. The experiment compared the performance of four methods, namely Binary Relevance, Classifier Chains, Label Power-set, and Fine-tuned BERT, and evaluated their effectiveness based on accuracy, F1-Score, Precision, and Recall values. The results revealed that the fine-tuned BERT method significantly outperformed the other three methods, achieving high accuracy, F1-Score, Precision, and Recall values. Binary Relevance also performed well, but Classifier Chains and Label Power-set exhibited relatively poor performance on this imbalanced dataset.

Overall, the findings suggest that finetuning the pre-trained BERT model is a good idea because it enables the model to adapt to a specific application. BERT is a powerful language model that has been pre-trained on a large corpus of text, allowing it to learn the intricacies of language and syntax. However, while pre-training provides a strong foundation, it does not necessarily optimize the model for specific tasks such as sentiment analysis, question answering, or text classification. Finetuning allows us to take the pre-trained BERT model and tailor it to our specific needs by training it on a smaller dataset that is specific to the task at hand.

Reproducing the results of a finetuned BERT model requires the same preprocessing, architecture, and hyperparameters used in the original experiment. To reproduce the results, the same evaluation metrics should be used, and the results should be compared to the original experiment. However, the dataset used for finetuning the BERT model is highly dependent on the application. Different tasks require different datasets, and the quality and size of the dataset can greatly affect the model's performance. A small, poorly labeled dataset may not provide enough information for the model to learn, while a large, well-labeled dataset may enable the model to generalize better. Therefore, it is essential to choose an appropriate dataset for the specific task at hand to achieve optimal results.

The primary objective of this paper was not to provide a step-by-step guide for reproducing a specific model. Instead, the paper aims to introduce and advocate the idea of fine-tuning BERT models on imbalanced datasets to achieve superior performance.

### Acknowledgements

The authors thank the French company FirstECO (https://www.firsteco.fr/) for providing the dataset, the French government for the plan France Relance funding, and Cyril Nguyen Van for his assistance.